\documentclass[prb, aps, twocolumn, showpacs]{revtex4}
\usepackage{amssymb}
\usepackage{amsmath}
\usepackage[dvips]{graphicx}
\usepackage[english]{babel}

\begin{document}

\title{Applicability of Bardeen-Cooper-Schrieffer theory to small-sized superconductors: role of Cooper-pair binding energy}
\author{W. V. Pogosov$^{1,2,*}$}
\affiliation{(1) Center for Fundamental and Applied
Research, N. L. Dukhov All-Russia Research Institute of Automatics,
Sushchevskaya 22, 127055 Moscow, Russia} \affiliation{(2) Institute for Theoretical and Applied
Electrodynamics, Russian Academy of Sciences, Izhorskaya 13, 125412,
Moscow, Russia} \affiliation{(*)
Corresponding author. Tel. +7-495-3625147; Fax +7-495-4842633;
e-mail: Walter.Pogosov@gmail.com}

\begin{abstract}
We analyze conditions of applicability of grand-canonical mean-field
Bardeen-Cooper-Schrieffer theory to the evaluation of an interaction
energy in the ground state of small-sized superconductors. We argue
that this theory fails to describe correctly an interaction energy,
when an average distance between energy levels near the Fermi energy
due to the size quantization becomes of the order of the single-pair
binding energy. In conventional superconductors, this quantity is
much smaller than the superconducting gap.
\end{abstract}

\author{}
\maketitle
\date{\today }

\section{Introduction}

It is known that Cooper pairing between electrons in conventional
superconductors is due to their interaction through phonons. Basic
properties of such superconductors can be described within the framework of
Bardeen-Cooper-Schrieffer (BCS) theory of superconductivity. In the
BCS theory \cite{BCS}, electron pairing is accounted for by the interaction
term in the Hamiltonian, which couples only electrons with up and
down spins on the same energy level (reduced BCS Hamiltonian).

Furthermore, the problem is solved approximately by
using a mean-field approach. There are two well-known realizations of
this approach in connection to the BCS theory.
The first one uses a variational ansatz for the many-body wave
function of the electron system \cite{BCS}. The second one is based on the
Bogoliubov canonical transformation \cite{Bogoliubov}. It turns out that both methods
give essentially the same results. In addition, in both approaches a
grand-canonical description (electron number is not fixed) is
encoded which becomes asymptotically exact in the macroscopic limit.

Thus, BCS theory contains two sources of inaccuracies, namely,
mean-field method to solve the many-body problem and the
grand-canonical description. It was shown that inaccuracies of the
first type become negligible in a macroscopic limit due to the very
peculiar form of the interaction term of the reduced BCS Hamiltonian
\cite{Bogoliubov1,Bardeen,Lieb} (however, see a recent study of
correlation functions in Ref. \cite{Boris}). Hence, mean-field BCS
results for the low-lying part of energy spectrum, including an
important expression for the superconducting gap, are asymptotically
exact in the thermodynamical limit
(corrections are proportional to the inverse number of particles).
Note that there also exist formulations of BCS theory with fixed number of particles
\cite{fixed}.

However, when the system size is decreasing, one can expect that
macroscopic BCS results at certain stage become no longer
applicable, since inaccuracies of both types start to be significant
(despite of the fact that the reduced BCS Hamiltonian itself is
adequate for the case of metallic grains \cite{Aleiner}).

A problem of applicability of the mean-field BCS treatment to
small-sized systems was addressed in literature for decades. A
widely spread point of view is due to the well-known paper of
Matveev and Larkin \cite{ML}, who concentrated on parity effects.
They introduced a parameter aimed to characterize these effects,
this quantity being defined as $\Delta _{ML} =
E_{2l+1}-\frac{1}{2}(E_{2l}+E_{2l+2})$, where $E_{N}$ is a ground
state energy as a function of the number of electrons. In the
macroscopic limit, $\Delta _{ML}$
reduces to the usual superconducting gap $\Delta$. Matveev and
Larkin have shown, by considering deviations from the mean-field
treatment that $\Delta _{ML}$ starts to differ from $\Delta $,
provided that a typical distance between one-electron energy levels
$\delta \varepsilon$ due to the size quantization becomes of the
order of $\Delta$. From this result it is usually concluded that BCS
theory is adequate until $\delta \varepsilon < \Delta$. Indeed,
$\Delta$ yields a natural energy scale associated with the Cooper
pairing and, therefore, such a conclusion is expectable.

A parity effect, as described by the Matveev-Larkin parameter
$\Delta _{ML}$, is an important characteristic of the interaction in
the Cooper channel. However, it is definitely not the only one
quantity which is fully determined by it. Another quite natural
quantity is an interaction energy in the ground state, when all
electrons are paired, as well as a closely related quantity, which
is a condensation energy. The interaction energy is an energy
difference between the ground states of the same electron system
without and with the interaction. For the condensation energy we use
a definition of Ref. \cite{Schechter}. Namely, it is the difference
between the expectation value of BCS Hamiltonian in the Fermi ground
state and the exact ground state energy. It is more difficult to
probe experimentally the ground state, but the condensation energy
influences, for example, spin magnetization of grains
\cite{Schechter}.

The evolution of the ground state energy upon changing system's size
was studied in Ref. \cite{Schechter} within the conserved particle
number approach and using the Richardson exact solution of the
reduced BCS Hamiltonian \cite{Rich1}. Quite surprisingly, it was
shown in Ref. \cite{Schechter} that the condensation energy starts
to deviate from the prediction of the grand-canonical BCS theory
much earlier compared to what could be expected from the
Matveev-Larkin criterion (see also Ref. \cite{Alt}). Namely, it
happens when $\delta \varepsilon $ becomes of the order of
$\Delta^{2}/\omega_{D}$, where $\omega_{D}$ is a Debye frequency.
Since in conventional superconductors $\Delta \ll \omega_{D}$, this
quantity is much smaller than $\Delta $. Let us stress that,
of course, this result does not contradict a general conclusion on
the asymptotical accuracy of the
mean-field BCS solution in the thermodynamical limit.
Nevertheless, the result of Ref. \cite{Schechter} seems to be rather puzzling, since it is believed that there is only a single energy scale characterizing superconducting
correlations which is a gap $\Delta $. It implies that there is
also another scale $\Delta^{2}/\omega_{D}$, but the physical meaning
of this quantity as well as its origin are not so clear.

In this paper, we present various arguments in favor of the point of
view that the additional energy scale reported in Ref.
\cite{Schechter} is nothing but a binding energy of a single pair
found by Cooper long time ago \cite{Cooper}. We point out that the
criterion of Ref. \cite{Schechter} is in an agreement with the
approximate expression for the ground state energy obtained recently
in Ref. \cite{EPJB}, where single-pair binding energy was used as a
'building block' for the total energy of the many-pair system. We
also demonstrate that a criterion involving $\Delta^{2}/\omega_{D}$
can be recovered using a perturbation theory or simple dimensional
estimates. A concept of an energy scale given by a binding energy of
a single Cooper pair was already introduced in Refs.
\cite{Monique,JPhys} in connection to the macroscopic BCS limit.
Single-pair binding energy provides a hidden energy scale which was
used to develop a novel interpretation of known BCS results
\cite{Monique}. Arguments of this paper suggest that this energy
scale becomes explicit in small-sized systems, where deviations from
the mean-field treatment are significant.

\section{Model}

We consider a system of finite sizes which contains
electrons with up and down spins. This system is described by the
reduced BCS Hamiltonian, $H=H_{0}+\mathcal{V}$. The first term,
$H_{0}$, is a kinetic energy
\begin{equation}
H_{0}=\sum_{n}\varepsilon _{n}\left(
a_{n\uparrow
}^{\dagger }a_{n\uparrow }+a_{n\downarrow }^{\dagger }a_{%
n\downarrow }\right),  \label{H0}
\end{equation}
where $a_{n\uparrow}^{\dagger }$ ($a_{n\downarrow }^{\dagger }$) is
a creation operator for
an electron in a discrete state $n$ with up
(down) spin. The second term, $\mathcal{V}$, is responsible for the
interaction between electrons with up and down spins:
\begin{equation}
\mathcal{V}=-V\sum_{n,n^{\prime
}}a_{n\uparrow }^{\dagger }a_{n\downarrow }^{\dagger }a_{%
n^{\prime }\downarrow }a_{n^{\prime }\uparrow }.  \label{BCSpot}
\end{equation}
This interaction acts only for the states within the 'Debye window',
$E_{F}-\omega_{D} < \varepsilon _{n} < E_{F}+\omega_{D}$. The
average spacing between neighboring energy levels within this
interval is $\delta \varepsilon $. We take into account finiteness
of the system size only by a nonzero value of $\delta \varepsilon $
which can be compatible to internal parameters characterizing
superconducting correlations (such as the gap). It is sufficient,
for the purposes of the present paper, to employ a so-called
equally-spaced model which assumes that energy levels are
distributed equidistantly within the Debye window. The density of
states $N(0)$ within this window is equal to $1/\delta \varepsilon$.
A dimensionless interaction constant $v \equiv N(0)V$ is a
characteristic of a given material and it can be taken nearly
independent on system size ($v$ is small for most conventional
superconductors). Thus, with the decrease of the system size,
$\delta \varepsilon $ increases, while $V$ decreases.

By using standard mean-field methods, it is possible to derive BCS
expression of the interaction energy. This quantity is extensive and
coincides with the condensation energy up to intensive terms. In the
macroscopic limit, the expression reads as
\begin{equation}
E_{int}^{(BCS)}\simeq \frac{1}{2}N(0)\Delta^{2}=\frac{1}{2}
\frac{\Delta^{2}}{\delta \varepsilon}, \label{BCScond}
\end{equation}
where $\Delta \simeq 2\omega_{D} \exp(-1/v)$. In this limit, $\delta
\varepsilon $ is the smallest energy scale in the system.

\section{Richardson solution and electron-hole symmetry}

Instead of using a mean-field approximation, one can solve the BCS
reduced Hamiltonian through the Richardson exact approach, which
assumes a fixed number of electrons. Within this approach the
problem is reduced to the resolution of the system of nonlinear
algebraic equations (Bethe equations) for the set of energy-like
quantities which determine a Hamiltonian eigen-state. Unfortunately,
very few controllable analytical solutions are known for these
equations \cite{Duk}. None of them covers an important crossover
regime between the two limits: $\delta \varepsilon \gg \Delta$ and
$\delta \varepsilon \ll \Delta$, where deviations from BCS results
are significant.

Quite recently, a simple universal formula for the ground state
energy was proposed in Ref. \cite{EPJB}. The derivation of Ref.
\cite{EPJB} is based on symmetry between electron pairs and hole
pairs for the reduced BCS Hamiltonian in the case of the
equally-spaced model. It also uses certain assumption on the
structure of the solution which is satisfied in all limits solvable
analytically. The expression of interaction energy can be written as
\begin{equation}
E_{int}^{(e-h)}=\frac{1}{2}\frac{\omega_{D}}{\delta
\varepsilon}(\varepsilon_{c}+V), \label{Interpcond}
\end{equation}
where $\varepsilon_{c}$ is given by the solution of a single
Richardson equation:
\begin{equation}
\frac{1}{V}=\sum_{n=0}^{2\omega_{D} / \delta \varepsilon}\frac{1}{2n
\delta \varepsilon + \varepsilon_{c}}, \label{Cooper}
\end{equation}
where $2\omega_{D} / \delta \varepsilon$ is the number of
one-electron energy levels in the Debye window. The latter equation,
which in general case must be solved numerically, can be rewritten
in terms of digamma-function:
\begin{equation}
\frac{2 \delta \varepsilon}{V}=\psi \left(\frac{2\omega_{D}}{\delta
\varepsilon}+\frac{\varepsilon_{c}}{2\delta \varepsilon}\right)-\psi
\left(\frac{ \varepsilon_{c} }{2 \delta \varepsilon}\right).
\label{digamma}
\end{equation}

It is easy to see that $\varepsilon_{c}$ is nothing but the binding
energy of a single pair, as introduced long time ago by Cooper
\cite{Cooper}. He considered a macroscopic system, $\omega_{D} /
\delta \varepsilon \rightarrow \infty$, which however accommodates
only a single pair in the Debye window, while in the BCS
configuration this window is half-filled and the number of electron
pairs in it,  $\omega_{D} / \delta \varepsilon$, is macroscopically
large.

Eqs. (\ref{Interpcond}) and (\ref{Cooper}) provide \cite{EPJB}
asymptotically exact results in the solvable limits $\delta
\varepsilon \gg \Delta$ and $\delta \varepsilon \ll \Delta$.
For instance, in the macroscopic limit, $\omega_{D} / \delta \varepsilon \rightarrow
\infty$, the summation in Eq. (\ref{Cooper}) can be replaced by the
integration, yielding $\varepsilon_{c}^{(macro)} \simeq 4\omega_{D}
\exp(-2/v)$. By substituting this expression to Eq.
(\ref{Interpcond}), we recover Eq. (\ref{BCScond}).
Moreover, the comparison with the results of the numerical solution
of the whole set of Richardson equations revealed a good accuracy of
this approach in the crossover region \cite{EPJB}.

Since Eqs. (\ref{Interpcond}) and (\ref{Cooper}) provide a good
approximation for the crossover regime and contain physically
meaningful quantities, one can answer a question under what
conditions Eq. (\ref{BCScond}) is no longer accurate. It is easy to
see from Eqs. (\ref{Cooper}) and (\ref{digamma}) that this happens
when it is not possible anymore to replace summation by integration
in Eq. (\ref{Cooper}). It follows directly from Eq. (\ref{digamma})
that the replacement is not allowed when $\delta \varepsilon \gtrsim
\varepsilon_{c}^{(macro)}$. This criterion for the interaction
energy differs significantly from the condition obtained in Ref.
\cite{ML} for $\Delta_{ML}$, because $\varepsilon_{c}^{(macro)} \ll
\Delta$ at $v$ small. Note that the expressions of
$\varepsilon_{c}^{(macro)}$ and $ \Delta$ look similarly, but these
quantities differ dramatically due to the difference in the
exponents.

Let us now discuss a connection of the obtained result with the
conclusion of Ref. \cite{Schechter}, where it was found that BCS
approach for condensation energy fails at $\delta \varepsilon \sim
\Delta^{2}/\omega_{D}$. Condensation energy differs from the
interaction energy only by a contribution $v \omega_{D}$ and
therefore the criteria of applicability of BCS theory to both these
quantities are expected to be quite similar. A direct substitution
of the expression of $\Delta$ shows that $\Delta^{2}/\omega_{D}$ is
nothing but $\varepsilon_{c}^{(macro)}$ in a full agreement with our
findings.

\section{Perturbation theory}

A similar criterion for the applicability of BCS theory to
interaction energy can be recovered without turning to the exact
Richardson solution, but using a perturbation theory. In order to
demonstrate it, let us split the interaction term $\mathcal{V}$ of
the Hamiltonian into a sum of two terms. The first term, $V_{0}$, is
chosen in such a way that BCS solution corresponds to the exact
eigen-states of $H_{0}+V_{0}$, while the second term, $\mathcal{W}$,
can be treated as a perturbation. Actually, such a splitting can be
performed by linearizing interaction term $\mathcal{V}$ of
Hamiltonian in deviations of products of two fermionic operators
from their mean values. The expressions of $V_{0}$ and $\mathcal{W}$
read as
\begin{equation}
V_{0}=\frac{\Delta^{2}}{V}-\Delta \sum_{n}(a_{n\uparrow} ^{\dagger
}a_{n\downarrow} ^{\dagger }+a_{n\downarrow }a_{n\uparrow }),
\label{V0}
\end{equation}
\begin{equation}
\mathcal{W}=-V \left( \sum_{n}a_{n\uparrow} ^{\dagger
}a_{n\downarrow} ^{\dagger }-\Delta/V \right) \left( \sum_{n^{\prime
}}a_{n^{\prime }\uparrow }a_{n^{\prime }\downarrow }-\Delta/V
\right),\label{W}
\end{equation}
while $\Delta$ is given by the usual condition
\begin{equation}
\Delta=V \sum_{n} \langle a_{n\uparrow} ^{\dagger }a_{n\downarrow}
^{\dagger } \rangle, \label{deltadefin}
\end{equation}%
which makes it evident that the grand-canonical description is
utilized within this approach.

It is easy to see that the interaction energy associated with
$H_{0}+V_{0}$ is given by Eq. (\ref{BCScond}). When deriving this
expression, one has to replace summation by the integration in the
gap equation as well as in the expression of the ground state
energy. This cannot be done if $\delta \varepsilon \gtrsim \Delta$.
Otherwise, finite-size corrections \emph{within} the mean-field
approximation become large. However, a perturbation $\mathcal{W}$
responsible for fluctuations leads to significant complications. The
effect of this perturbation in the thermodynamical limit was
recently discussed in Ref. \cite{WePhysC}. The same technique can be
applied for the system with finite $\delta \varepsilon $. Using Eq.
(69) of Ref. \cite{WePhysC}, we readily find a first-order
correction in $\mathcal{W}$ to the interaction energy as $v
\omega_{D}$. It is of the same sign as the dominant contribution,
i.e., superconducting correlations are enhanced by taking into
account corrections \emph{beyond} the mean-field approximation.

We now compare the dominant contribution to the interaction energy
given by Eq. (\ref{BCScond}) with the first-order correction in
$\mathcal{W}$. We find that they become of the same order at
\begin{equation}
\delta \varepsilon \sim \frac{\Delta^{2}}{\omega_{D}} \ln
\frac{2\omega_{D}}{\Delta}. \label{pert-cond}
\end{equation}%
This quantity is much smaller than $\Delta$. Moreover, in the view
of the relation $\Delta \ll \omega_{D}$ at small $v$, we can omit a
weak logarithmic factor in the right-hand side of Eq.
(\ref{pert-cond}). Hence, we again arrive to the same criterion.
Note that the first order correction in $\mathcal{W}$ to the
condensation energy vanishes. However, one can estimate a
second-order correction in $\mathcal{W}$ using Eq. (77) of Ref.
\cite{WePhysC} as $\sim v^{2} \omega_{D}$. This term does contribute
to condensation energy and again supports the criterion
suggested in Ref. \cite{Schechter}.

\section{Dimensional estimates}

Perhaps, the most elementary argument in favor of the discussed
criterion can be suggested by applying a method of dimensional
estimates. Namely, we know that in the limit $\delta \varepsilon \ll
\Delta$ the interaction energy is given by Eq. (\ref{BCScond}). In
the opposite limit $\delta \varepsilon \gg \Delta$ it is given again
by $v \omega_{D}$, as can be readily obtained by constructing a
perturbation expansion in $\mathcal{V}$ around $H_{0}$, i.e.,
starting from the system of normal electrons (the same result can be
easily recovered using Richardson approach). Comparing expressions
for the interaction energy in both limits, we immediately arrive at
Eq. (\ref{pert-cond}).

\section{Discussion and summary}

Thus, a puzzling parameter $\Delta^{2}/\omega_{D}$ turns to be
nothing but the binding energy of a single pair. The existence
of this quantity in connection to \emph{macroscopic} results of BCS
theory was recently discussed in Refs. \cite{Monique,JPhys}.
Actually, based on this quantity, it is possible to develop an
interesting interpretation of usual BCS expression (\ref{BCScond})
for the interaction energy. Standard understanding uses a fact that
the interaction between the electrons smears out Fermi energy of
noninteracting electrons only within a narrow layer having a width
$\sim \Delta$, while the number of such 'excited' pairs is $\sim
\Delta / \delta \varepsilon$. By ascribing an energy $\sim \Delta $
to each pair, one recovers Eq. (\ref{BCScond}) for the interaction
energy.

The understanding proposed in Refs. \cite{Monique,JPhys} relies
on the fact that all electrons within the Debye window are subjected
into the interaction, so that the total number of such pairs is
$\omega_{D} / \delta \varepsilon$. This is also consistent with the
form of the BCS wave function \cite{BCS}. The binding energy of each
pair is given by a single-pair binding energy
$\varepsilon_{c}^{(macro)}$ reduced by the factor of 2. The
reduction is due to the fact that the number of unoccupied
electronic states available for the formation of a pair is also
twice smaller compared to the Cooper configuration. We thus again
recover Eq. (\ref{BCScond}), but without using an energy scale of $
\Delta $. Qualitatively, this understanding is deeply linked to the
mean-field concept of each pair being immersed into an average field
due to remaining \emph{identical} pairs. The existence of the energy
scale $\varepsilon_{c}^{(macro)}$ is thus also connected to the
validity of the mean-field picture in the macroscopic limit.

The fact that in the many-pair system we still see appearing a
single-pair binding energy should be related to the very peculiar
form of the BCS reduced Hamiltonian, which contains only coupling
between up and down spin electrons on the same energy level, so that
pairs feel each other only kinematically, i.e., through the Pauli exclusion principle for
the constituent electrons, and not dynamically.

It is easy to realize that although this picture provides new
insights to the problem, it does not predict any consequences for
measurable quantities in the macroscopic limit at zero temperature,
when the mean-field treatment is asymptotically exact. However, as
follows from the results of this paper, a hidden energy scale
$\varepsilon_{c}^{(macro)}$, associated with the validity of the
mean-field picture, becomes explicit in small-sized systems,
\emph{when deviations from the mean-field BCS predictions are not
negligible anymore}. From this viewpoint, the major result of the
present paper is rather natural.

Let us stress that the discussed criterion is valid only for those quantities,
which are determined by the condensation energy. For those characteristics, which are determined by excitation energies and pair-breaking effects (such as critical temperature $T_{c}$), the criterion of Ref. \cite{ML} must be adequate.

We would like also to mention that, in this paper, we considered the simplest possible model, which enables us to isolate the studied effect from many other important contributions inevitable in real experiments. Among them one can mention level broadening \cite{Bose}, mesoscopic fluctuations \cite{Dussel,Alhassid}, and spatially nonuniform pairing \cite{Croitoru1,Croitoru2}. The effect of level broadening, however, is more important for temperatures close to $T_{c}$, while we here consider the low-temperature regime \cite{Bose}. In addition, it is also stronger for smaller systems, in which the mean interlevel spacing is of the order of the superconducting gap. This latter circumstance also applies for mesoscopic fluctuations in the case of disordered grains. For such grains, it was demonstrated \cite{Dussel} that the crossover between the superconducting state and the fluctuation-dominated regime is
smooth, similarly to the case of the equally-spaced model. Therefore, the discussed criterion must be valid for such grains, but one should use a mean interlevel spacing in it. This has to be contrasted with nano-superconductors of highly-symmetric shapes in a clean limit, for which shell effects as well as a spatially nonuniform pairing take place\cite{Croitoru1,Croitoru2}. However, these effects are again more pronounced for smaller systems. Nevertheless, it is also perspective to study in a more detail an accuracy of the mean-field theory for the evaluation of various superconducting characteristics of such systems.

In summary, we discussed a criterion of applicability of the
grand-canonical mean-field BCS theory to the evaluation of
interaction energy in the ground state of a small-sized system. Based
on several approaches and arguments, we demonstrated that
the result of this theory for the interaction energy is accurate unless
the average difference between the
energies of one-electron states becomes of the order of the binding
energy of a single Cooper pair. This quantity is much smaller than
the superconducting gap.

\acknowledgments

Useful discussions with M. Combescot are acknowledged.
This work was supported by the RFBR (project no. 12-02-00339),
RFBR-CNRS programme (project no. 12-02-91055), and
Russian Science Support Foundation.


\begin{references}

\bibitem{BCS}J. Bardeen, L. N. Cooper, and J. R. Schrieffer, Phys. Rev. {\bf 108}, 1175
(1957).

\bibitem{Bogoliubov}N. N. Bogoliubov, ZhETF {\bf 34}, 58 (1958)  [Sov. Phys. JETP {\bf 7}, 41 (1958)];
N. N. Bogoliubov, Nuovo Cimento {\bf 7}, 794 (1958).

\bibitem{Bogoliubov1}N. N. Bogoliubov, Physica {\bf 26}, S1 (1960).

\bibitem{Bardeen}J. Bardeen and G. Rickayzen, Phys. Rev. {\bf 118}, 936 (1960).

\bibitem{Lieb}D. Mattis and E. Lieb, J. Math. Phys. {\bf 2}, 602 (1961).

\bibitem{Boris}O. El Araby and D. Baeriswyl, Phys. Rev. B \textbf{89}, 134521 (2014).

\bibitem{fixed}K. Dietrich, H. J. Mang, and J. H. Pradal, Phys. Rev. B \textbf{22}, 135 (1964);
F. Braun and J. von Delft, Phys. Rev. Lett. \textbf{81}, 4712
(1998).

\bibitem{Aleiner}I. L. Kurland, I. L. Aleiner, and B. L. Altshuler, Phys. Rev. B \textbf{62}, 14886 (2000).

\bibitem{ML}K. A. Matveev and A. L. Larkin, Phys. Rev. Lett. \textbf{78}, 3749 (1997).

\bibitem{Schechter}M. Schechter, Y. Imry, Y. Levinson, and J. von Delft, Phys. Rev. B \textbf{63}, 214518 (2001).

\bibitem{Rich1}R. W. Richardson, Phys. Lett. {\bf 3}, 277 (1963).

\bibitem{Alt}E. A. Yuzbashyan, A. A. Baytin, and B. L. Altshuler,
Phys. Rev. B {\bf 71}, 094505 (2005).

\bibitem{Cooper}L. N. Cooper, Phys. Rev. {\bf 104}, 1189 (1956).

\bibitem{EPJB}W. V. Pogosov, N. S. Lin, and V. R. Misko, Eur. Phys. J. B \textbf{86}, 235 (2013).

\bibitem{WePhysC}M. Combescot, W. V. Pogosov, and O. Betbeder-Matibet, Physica C \textbf{485}, 47 (2013).

\bibitem{Monique}W. V. Pogosov and M. Combescot, Pis'ma
v ZhETF {\bf 92}, 534  (2010) [JETP Letters {\bf 92}, 534 (2010)];
M. Crouzeix and M. Combescot, Phys. Rev. Lett. \textbf{107}, 267001 (2011).

\bibitem{JPhys}W. V. Pogosov, J. Phys.: Condens. Matter \textbf{24}, 075701 (2012).

\bibitem{Duk}J. Dukelsky, S. Pittel, and G. Sierra, Rev. Mod. Phys. {\bf 76}, 643 (2004).

\bibitem{Bose}I. Brihuega, A. M. Garc\'ia-Garc\'ia, P. Ribeiro, M. M. Ugeda, C. H. Michaelis, S. Bose, and K. Kern, Phys. Rev. B \textbf{84}, 219906 (2011).

\bibitem{Dussel}G. Sierra, J. Dukelsky, G. G. Dussel, J. von Delft, and F. Braun, Phys. Rev. B \textbf{61}, R11890 (2000).

\bibitem{Alhassid}K. N. Nesterov and Y. Alhassid, Phys. Rev. B \textbf{87}, 014515 (2013).

\bibitem{Croitoru1}M. D. Croitoru, A. A. Shanenko, C. C. Kaun, and F. M. Peeters, Phys. Rev. B \textbf{83}, 214509 (2011).

\bibitem{Croitoru2}M. D. Croitoru, A. A. Shanenko, F. M. Peeters, and V. M. Axt, Phys. Rev. B \textbf{84}, 214518 (2011).

\end{references}
\end{document}